\documentclass{elsart}

\usepackage{harvard}

 \usepackage{graphics}

\usepackage{amssymb}

\def\astrobj#1{#1}
\def\url#1{{\ttfamily\def\/{/\discretionary{}{}{}}#1}}

\newcommand{\til}{$\sim$}
\def\spose#1{\hbox to 0pt{#1\hss}}
\def\simlt{\mathrel{\spose{\lower 3pt\hbox{$\mathchar"218$}}
     \raise 2.0pt\hbox{$\mathchar"13C$}}}
\def\simgt{\mathrel{\spose{\lower 3pt\hbox{$\mathchar"218$}}
     \raise 2.0pt\hbox{$\mathchar"13E$}}}
\def\arcsec{\nobreak\ifmmode{''}\else{$''$}\fi}
\def\degree{\nobreak\ifmmode{^\circ}\else{$^\circ$}\fi}
\newcommand{\ergsec}{\thinspace\hbox{$\hbox{erg}\thinspace\hbox{s}^{-1}$}}
\newcommand{\msun}{\thinspace\hbox{$M_{\odot}$}}
\newcommand{\rsun}{\thinspace\hbox{$R_{\odot}$}}
\newcommand{\lsun}{\thinspace\hbox{$L_{\odot}$}}
\newcommand{\I}{{\scriptsize~I}}

\begin{document}
\normalsize{

\begin{frontmatter}
\title{Period derivative of the M15 X-ray Binary AC211/X2127+119}
\author{L. Homer\thanksref{email1}} and
\author{P. A. Charles\thanksref{email2}}
\address{Department of Astrophysics, Nuclear Physics Lab., Keble Road, Oxford OX1 3RH}

\begin{abstract}
We have combined {\it Rossi X-ray Timing Explorer} observations of X2127+119, the low-mass X-ray binary in the globular cluster \astrobj{M15}, with archival X-ray
lightcurves to study the stability of the 17.1 hr orbital period.  We find that the data cannot be fit by the \citeasnoun{ilov93} ephemeris,
and requires either a $7\sigma$ change to the period or a period derivative $\dot{P}/P\sim9\times10^{-7} {\rm yr^{-1}}$.  Given its remarkably low $L_X/L_{opt}$ such
a $\dot{P}$ lends support to models that require super-Eddington mass transfer in a $q\sim1$ binary.
\end{abstract}

\thanks[email1]{lh@astro.ox.ac.uk}
\thanks[email2]{pac@astro.ox.ac.uk}
\end{frontmatter}

\section{Introduction}

The X-ray source \astrobj{X2127+119} is one of the \til10 bright ($>10^{36}\ergsec$) low mass X-ray binaries (LMXBs) located within globular clusters, being within $2\arcsec$ of
the core of \astrobj{M15}.   It was the first cluster X-ray source to have an optical counterpart proposed, the V\til15 star \astrobj{AC211} \cite{auri84}, on the basis of blue colours, variability and location within the {\it Einstein} HRI error
box \cite{geff89}.  The identification was spectroscopically confirmed by \citeasnoun{char86},
who found He{\scriptsize~II} $\lambda4686$ and H$\alpha$ emission.  \astrobj{AC211} is optically one of the brightest of all LMXBs, which is remarkable given its relatively low $L_X$.  Consequently, it has been extensively studied \cite{hert87,call88,dota90,ilov87,nayl88,bail89,ilov93} both in X-rays and at optical/UV wavelengths.  However, deriving the system parameters and understanding
the geometry has proven difficult.  Observations have yielded a wide variety of temporal behaviour, obscuring the true binary period, and other
unusual phenomena, as detailed below. 

\astrobj{AC211} exhibits the largest amplitude optical variations of any LMXB, and \til5 times greater than in X-rays.  Following their first photometric campaign \citeasnoun{ilov87}
detected a 8.54hr periodic modulation in their series of U-band CCD images,
although there were puzzling unmodulated episodes. A reanalysis of  {\it HEAO-1} scanning data from November 1977 by \citeasnoun{hert87}
revealed a consistent 8.66hr X-ray modulation.  However, further optical spectroscopy by \citeasnoun{nayl88} gave a slightly different binary period of $9.1\pm0.5$hr
based on a radial velocity study of the He{\scriptsize~I} $\lambda4471$ absorption line.  More remarkably, they found that the
$\gamma$ velocity was
blue-shifted by \til$150\pm10 \rm{km s}^{-1}$ with respect to the cluster.  Their favoured model for the system consisted of a high-inclination
accretion disc corona source, which accounts for the unusually low $L_X/L_{opt}\sim20$.   A variable height disc rim (greatest when close to the
accretion stream impact point) would then modulate both the X-rays and optical emission from the X-ray irradiated disc.  They invoked ejection
of the system from the cluster core (following a close encounter with another star) to explain the high $\gamma$ velocity. Alternative
explanations were suggested;
\citeasnoun{fabi87} proposed He{\scriptsize~I} absorption in a supersonic wind from a massive corona, whilst \citeasnoun{bail89} preferred an
outflow from the outer Lagrangian point as the site, requiring a common envelope system (the result of unstable mass transfer).  The detection of a radius expansion X-ray burst with {\it
Ginga} \cite{dota90} further complicated the picture, as clearly, in spite of the corona, the X-ray source must be directly visible for at least part of the time. Notably, this remains one of the most luminous X-ray bursts ever detected.

The most recent piece in the puzzle arose from the continuation of Ilovaisky and co-workers' photometric campaign.   \citeasnoun{ilov93}
(hereafter Il93)
analysed all their U-band data from 1984-89 as a single set, using a phase dispersion minimisation (PDM) search method \cite{stell78}.  This
yielded a preferred period of
17.11 hr, double that previously accepted.  This was able to explain the strange unmodulated observations and was found to be consistent with
all the X-ray data up to that date.
Using the minima in their lightcurve combined with that of the 1977  {\it HEAO-1} data gave an ephemeris of $\phi=0.0$ for JD
$2447790.963\pm0.018+n\times(0.713014{\rm d}\pm0.000001{\rm d})$.

Nevertheless, the \astrobj{AC211} system remains an enigma.  A high mass transfer rate, leading to significant mass loss from the system is a
distinct possibility.  Important constraints can therefore be placed by the measurement of any corresponding period change. Now that 9 years
have elapsed since the last period and ephemeris determination we decided to make use of the opportunity presented by over 2 years of {\it
Rossi X-ray Timing Explorer} All SKY Monitor ({\it 
RXTE}/ASM) observations to revisit \astrobj{AC211}.

\section{Determination of the revised ephemeris}
\subsection{Data acquisition}
The HEASARC maintains comprehensive databases of past and present X-ray missions, from which we obtained the {\it EXOSAT}, {\it ASCA}/SIS
and GIS lightcurves and the definitive dwell {\it RXTE}/ASM data on
\astrobj{X2127+119}.  Unfortunately, the {\it HEAO-1} and {\it Ginga} products are not readily usable, hence
we simply digitised the data presented in the published lightcurve plots of \citeasnoun{hert87} and \citeasnoun{vP90} (their Fig. 1 in
both cases).  Lastly, the CCD
U-band data were only available in their published form, so we digitised the folded data as presented in Fig.8 of Il93. In spite
of the relatively
poor temporal sampling of the {\it RXTE}/ASM dwell data, we found that the extensive coverage provided good phase sampling if taken over a sufficiently
long time period.  We chose to consider these data in two halves (as divided by time), hereafter referred to as ASM1 and ASM2.  A log of all
the data used is given in Table \ref{tab:log}.

\begin{table}
\caption{A Journal of the X-ray/optical observations of \astrobj{X2127+119}/\astrobj{AC211}.}\label{tab:log}
\small{
\begin{tabular}{c c c c c c c c} \hline
Observatory&Detector&\multicolumn{2}{c}{Date (UT)} & Time  & Energy  & $L_X.\left(\frac{{\rm d}}{10\thinspace{\rm kpc}}\right)^2$ \\
				&			&Start	 	&End				&	span			& Range (keV)			&($10^{36}\ergsec$)\\
\hline
{\it HEAO-1}& A-1		&19.11.77& 23.11.77		&	96.4 hr					 				&	0.25-25				& \til9.5\\	
{\it EXOSAT}& ME		&30.06.84&	30.06.88			&	11.48 hr					 				& 0.9-9		&	\til 8 \\
				&			&22.10.84&22.10.84		&	15.27 hr					 				&				&	\til 8 \\
				&			&20.10.85&20.10.85		&	7.17 hr					 				&				&	\til 8 \\
{\it Ginga}& LAC		&20.10.88&24.10.88		&	84.0 hr					 				&1-28			&	6.4	\\
{\it ASCA}& SIS		&16.05.95&17.05.95		&	26.3 hr					 				&0.4-12	&	4.2	\\
{\it ASCA}& GIS		&16.05.95&17.05.95		&	26.3 hr					 				&0.7-10	&	4.2	\\
{\it RXTE} &ASM		&23.02.96		&14.03.97						&	384.6 d					 				&	2-10				& 4.2			\\
&		&14.03.97			&02.04.98						&	384.5 d					 				&				& 3.4			\\
				&			&		   &						&						  				&				&				\\

Optical&	CCD	&28.07.84&22.09.89& 		 1881.4 d	& U-band		&- \\
\end{tabular}
}
\vskip 5pt

\end{table}

\subsection{Measurement of the ephemerides}
All the data were folded on the ephemeris and period given in Il93, and binned, in order to increase the signal-to-noise. We used 40 phase
bins, except for the {\it HEAO-1} data (where we used 20), with each bin covering 0.05 (or 0.1) in phase and hence each overlapping the adjacent
bins by $\pm0.025\;\;(\pm0.05)$.  This provided both good phase resolution,
and some smoothing to reduce noise.  These folded lightcurves are presented in Fig. \ref{fig:flc}.  
\begin{center}	
\begin{figure*}
\resizebox*{1.05\textwidth}{0.55\textheight}{\rotatebox{0}{\includegraphics{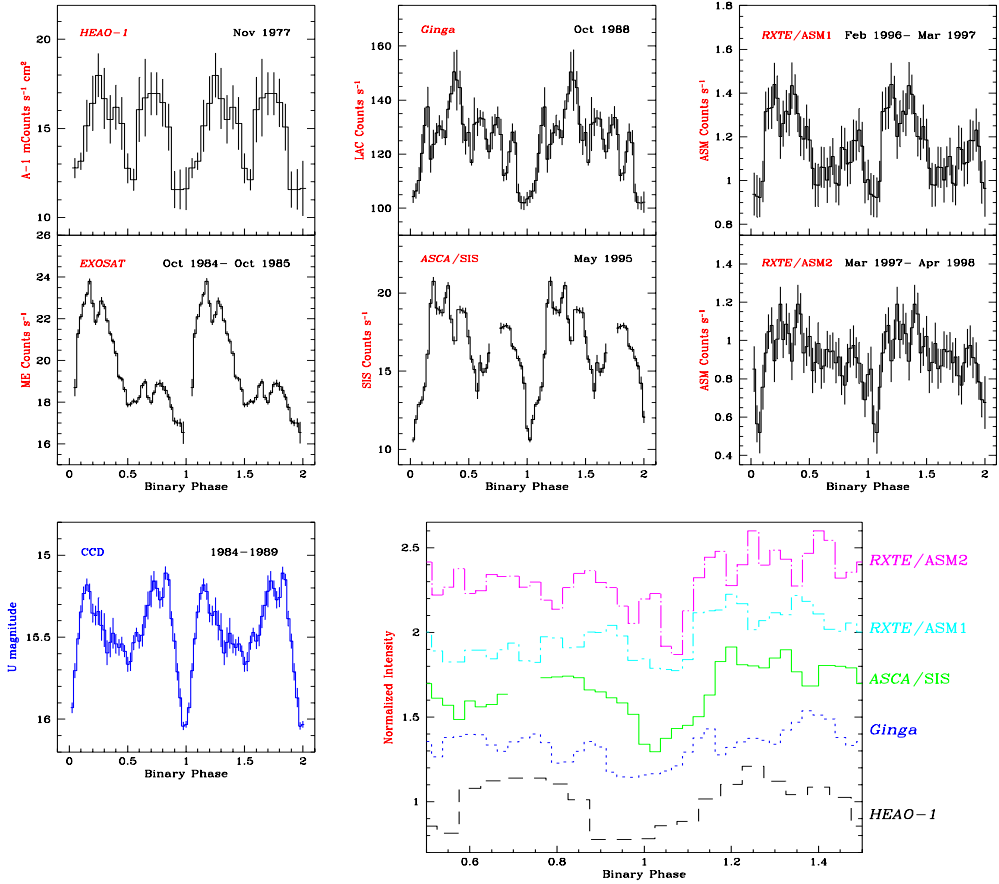}}}
\caption{Upper and lowest left panels: X-ray and optical lightcurves of X2127+119/AC211 respectively, folded (and binned)
on the Il93 ephemeris and period of 0.713014 d. Lowest right panel:  Sample of the X-ray lightcurves overlaid in chronological order (earliest lowest), with the flux units normalised and a vertical offset of 0.26 introduced between
each dataset for clarity.  Note the clear change of minimum phase over time.}\label{fig:flc}
\end{figure*}
\end{center}
As expected, given its probably different physical origins, the optical lightcurve (labelled CCD) is
different in shape to the X-ray, although both clearly demonstrate the eclipse by the secondary star.  However, it is clear that the X-ray lightcurves also show considerable
morphological evolution over time.  The energy ranges (see Table \ref{tab:log}) are in fact similar for all the X-ray observations and so the
changes must be intrinsic to the source.  Whilst the {\it EXOSAT} and ASM data are similar, the intervening  {\it Ginga} observation shows a totally different
shape.  

In order to determine the relative orbital ephemerides for each observation, we adopted the same approach as Il93.  They found the
extended eclipse, or minimum, to be the only stable feature of the lightcurve, and so we used a simple Gaussian fit to determine the time of minimum in each case. Due to
the poor phase coverage of the minimum in the 
{\it EXOSAT} data, we were forced to base our estimate on the sharp egress from minimum flux.  The results are plotted in the form of an O-C plot (i.e.
the phase offset between the expected minimum and that observed, versus cycle number) in the upper panel of Fig.\ref{fig:OC1}.
As a check, we also performed a cross correlation of the
{\it EXOSAT}, ASM1 and ASM2 folds with the complete ASM lightcurve fold to determine their relative
phase offsets (see lower panel of Fig.\ref{fig:OC1}; the values have been offset,
such that the mean ASM results agree with those of the upper panel). 
\begin{figure*}
\resizebox*{1.02\textwidth}{0.46\textheight}{\rotatebox{-90}{\includegraphics{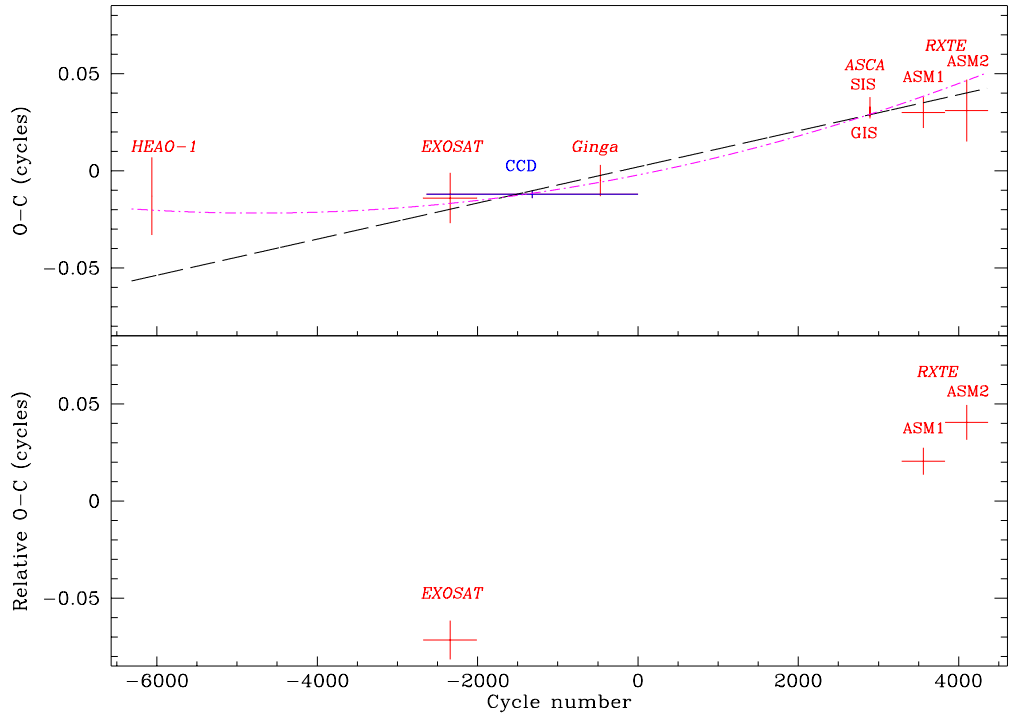}}}
\caption{O-C plot for the orbital ephemerides of \astrobj{X2127+119}/\astrobj{AC211} for each X-ray observation and the one optical dataset; as determined by a
Gaussian fit to each minimum (upper panel), and cross-correlations (lower panel).} \label{fig:OC1}
\end{figure*}

\subsection{Evidence for a period derivative}

A variety of model fits have been made to the O-C data: (i) a constant O-C corresponding to a linear ephemeris with the Il93 period of
0.713014 d is a very poor fit with $\chi^2_{\nu}=27.1$; (ii) a linearly increasing O-C corresponding to a linear ephemeris but with a different
period
of 0.713021\thinspace d $(7\sigma$ larger then Il93) is statistically a good fit with $\chi^2_{\nu}=1.12$; (iii) a parabolic O-C corresponding to an
ephemeris with the Il93 period and a period derivative of 
$\dot{P}=1.75\pm 0.90 \times10^{-9}$ is also a good fit with $\chi^2_{\nu}=0.59$.  However, one may apply a one-sided {\it F}-test
\cite{bevi92}, to test the improvement of fit for an additional polynomial term between models (ii) and (iii). $F_\chi=\Delta\chi^2/\chi^2_\nu=(6.752-2.954)/0.59=6.43$
corresponding to a 95\% confidence level that the parabolic fit is better than the linear one.  Moreover, the hypothesis of a linear
ephemeris is undermined by the need for a $7\sigma$ increase in the period, which is apparent in the discrepancy between the model and the {\it HEAO-1} point.
Comparison with the cross correlation results (lower panel) confirms that the data
require a significant change to the Il93 ephemeris, whilst demonstrating that the ephemeris change derived from the minima times is if anything
on the conservative side.    We therefore consider that the evidence points towards a constant period
derivative, with $\dot{P}/P
\sim 9\times 10^{-7} {\rm yr^{-1}}$. 
 

\section{Discussion}
\subsection{Physical origins of the X-ray and optical modulations}
\begin{center}
\begin{figure*}
\resizebox*{1.0\textwidth}{!}{\rotatebox{-90}{\includegraphics{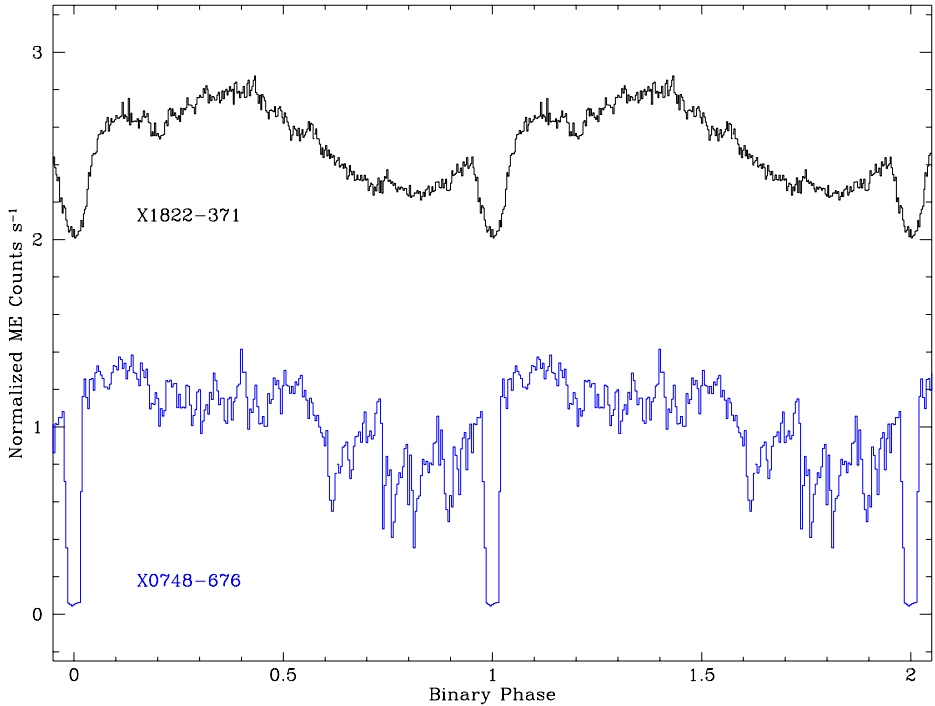}}}
\caption{The folded 1-10 keV {\it EXOSAT}/ME lightcurves of X1822-371 and X0748-676, using the period and ephemerides as given in White et
al. (1981) and Parmar et al. (1986) respectively.  The count rates have been normalized and the curves offset by 1 for clarity. }\label{fig:adc}
\end{figure*}  
\end{center}It is profitable to compare the lightcurve of \astrobj{X2127+119}/\astrobj{AC211} with those of the dipping sources X1822-371 and X0748-676 \cite{parm86}, as shown
in Fig.\ref{fig:adc}.  The X-ray lightcurve of X1822-371 clearly shows a broad minimum preceding, by 0.25 in phase, a deeper
partial eclipse by the companion star, which is coincident with
the optical minimum.  The
eclipse is only partial as the central source of this high inclination system is hidden by the disc and the X-ray scattering ADC is never fully
occulted.  To model the broad dip, \citeasnoun{whit82} required a thickened region of the disc $\sim100\degree$ upstream of the line
joining the companion and compact object.  X0748-676 is essentially very similar, but has a lower inclination, thereby enabling the central source to be
seen.  Here, the almost total X-ray eclipse is always preceded by complex dip structure, which varies from cycle to cycle, although there
appears to be a stable dip present at about $\phi=0.65$.  This latter feature is once again probably due to a thickened region of the disc.  Considering
the lightcurves of \astrobj{X2127+119}/\astrobj{AC211} in Fig.\ref{fig:flc}, the primary minima of every X-ray dataset are coincident with the optical  minimum,
but intervals of lower flux of variable duration are visible during phases  $\phi=0.4-1.0$.  This varying dip structure is certainly similar to
X0748-676.  The origin of the raised rim of the accretion disc is most probably  the impact 
of the accretion stream.  The understanding of this process remains incomplete, with a number of possible results from simulations, as outlined
in \citeasnoun{armi98} and references therein. In addition to a bulge at $\phi=0.8$, where the initial impact occurs, a reimpact following
deflection and then reconvergence of the stream at $\phi=0.6$ has been seen \cite{lubo89}, but the azimuthal extent of these regions is unclear.

If this interpretation is correct, then care clearly must be taken when determining the ephemeris of the binary motion from the times of primary
minima, as the dip structure will contribute to a varying degree to the flux reduction. This may well account for the changes seen in the asymmetry
of the minima. Further observations, in order to  both increase the base line of the ephemeris and to better model the details of the X-ray
modulation are obviously needed to more accurately measure the period derivative.

\subsection{Inferred mass transfer rate}
For this last section we will assume that $\dot{P}$ is within at least an order of magnitude of our estimate. The proposed
evolutionary scenario for \astrobj{AC211} involves a neutron star capturing a star close to the main sequence turn-off point in \astrobj{M15}, which soon evolves
towards the giant branch and
turns on mass transfer by overflowing its Roche lobe \cite{bail87}. Hence, one may apply the properties of a stripped giant to the secondary, which
depend only on the mass of its helium core $M_c$ and not its total mass $M_2$. 

The secondary radius $R_2$ (in $\rsun$) and luminosity $L_2$ (in $\lsun$) are given by the relations \cite{king88}:
\begin{equation}\label{eqn:rad} 
R_2=12.55\left(\frac{M_c}{0.25}\right)^{5.1}\, ,
\end{equation}
\begin{equation}\label{eqn:lum} 
L_2=33\left(\frac{M_c}{0.25}\right)^{8.11}
\end{equation}where $M_c$ is in $\msun$ and similarly for $M_2$ and $M_X$ below.
Using the \citeasnoun{pacz71} approximation for the Roche lobe radius:
\begin{equation}\label{eqn:RL} 
R_L=0.462a\left(\frac{M_2}{M_2+M_X}\right)^{1/3}
\end{equation}
then eliminating the binary separation $a$ by use of Kepler's law and applying $R_2=R_L$ for the mass transfer condition yields:
\begin{equation}\label{eqn:mmp}  
\left(\frac{M_c}{0.25}\right)^{7.65} M_2^{-0.5}=0.0609 \left(\frac{P}{\rm d}\right)
\end{equation}Taking the time derivative and dividing by (\ref{eqn:mmp}):
\begin{equation}\label{eqn:mmpr}  
-0.5\frac{\dot{M_2}}{M_2}+7.65\frac{\dot{M_c}}{M_c}=\frac{\dot{P}}{P}
\end{equation}The rate of change of $M_c$ with respect to $M_2$ is governed by nuclear evolution and for $q=M_2/M_x<\frac{5}{6}$ \cite{king88}:
\begin{equation}\label{eqn:mmpr}  
-\frac{\dot{M_2}}{M_2}=\frac{5.1}{\frac{5}{3}-2q}\frac{\dot{M_c}}{M_c}
\end{equation}Hence, we need to estimate the mass ratio $q$.  There are two limiting cases for $M_2$, when $M_c=M_2$ and $M_c=0.17M_2$.  The former is obvious whilst the latter follows from the need for $M_c/M_2
\simgt0.17$, which is the Sch\"{o}nberg-Chandrasekhar limit, if the secondary is to have left
 the main sequence.  Substitution into equation (\ref{eqn:mmp}), with the 0.713014 d period gives $M_{2,min}=0.146\msun$ and $M_{2,max}=0.92\msun$.  Thus
the range of masses given in Il93 of $0.74-0.81\msun$ (based upon the \astrobj{M15} isochrones and the appropriate age) fit well within these limits,
towards the upper end.  

Finally, taking $M_2=0.8\msun$ as a representative value and a canonical $1.4\msun$ neutron star ($q=0.57$) yields:
\begin{equation}
-\frac{\dot{M_2}}{M_2}=9.7\frac{\dot{M_c}}{M_c}\, ,
\end{equation}
\begin{equation} \label{eqn:mtr} 
\dot{M_2}=-0.77 M_2\frac{\dot{P}}{P}
\end{equation}and a mass transfer rate $\dot{M_2}\sim-6\times10^{-7} \msun {\rm yr}^{-1}$. 

The corresponding accretion luminosity is given by:

\begin{equation}\label{eqn:lacc} 
L_{acc}=-\eta\left(\frac{GM_X\dot{M_2}}{R_X}\right)=-\eta1.17\times10^{46}\left(\frac{\dot{M_2}}{\msun{\rm yr}^{-1}}\right)\;\; \ergsec 
\end{equation}

\[L_{acc}\sim\eta\, .\,6\times10^{39}\ergsec\ {\rm for}\;\; \dot{M_2}\sim-6\times10^{-7} \msun {\rm yr}^{-1}\]

The mass transfer rate is clearly super Eddington ($L_{E}\sim\eta\, .\,10^{38}\ergsec$ for a neutron star accreting hydrogen) and hence the
accretion efficiency $\eta\ll1$.  One would then expect significant mass loss from the system, after formation of a common
envelope.  This is just the scenario envisaged by \citeasnoun{bail89}, and indeed our estimate for $\dot{M_2}$ above is comparable to theirs,
which was based on the He\I\ absorption features.

\section{Conclusions}
We have obtained X-ray and optical lightcurves of X2127+119/AC211, from the HEASARC archives and published data, which span a total of 21
years, from  1977 {\it HEAO-1} data up to the most recent 1998 {\it RXTE}/ASM data.  Folding the lightcurves on the Il93 ephemeris and 17.1 hr
orbital period, we have examined the stability of this period.  The lightcurves in Fig.~\ref{fig:flc} clearly demonstrate the variety of X-ray morphologies, most probably arising from changes in the vertical structure of the outer edge of the accretion disc, which is responsible for
obscuring the X-ray emitting central region.  As a result caution is needed when using the folded lightcurves to check for any phase changes.  However, the principal minimum (due to
occultation by the secondary star) is obviously a persistent feature, and we have followed the procedure of Il93, to measure the relative
ephemerides of each curve, based on its phasing.  Only a non-zero
period derivative applied to the original period, provides a good fit to these data, implying $\dot{P}/P
\sim 9\times 10^{-7} {\rm yr^{-1}}$.  Modelling the secondary as an evolved sub-giant, it is possible to estimate the mass transfer rate necessary
to drive such a period change.  This yields $\dot{M_2}\sim-6\times10^{-7} \msun {\rm yr}^{-1}$, and hence an accretion luminosity
$L_{acc}\sim\eta\, .\,6\times10^{39}\ergsec\ $.  Lastly, as this far exceeds the
Eddington luminosity, the accretion efficiency must be very low , requiring there to be significant mass loss from the system.  This is
consistent with the earlier 
spectroscopic data of \citeasnoun{bail89}, and a $q\sim1$ binary.
 
\begin{ack}
This research has made use of data obtained through the High Energy Astrophysics Science Archive Research Center Online Service, provided by
the NASA/Goddard Space Flight Center.  In particular, we thank the ASM/RXTE teams at MIT and at the RXTE SOF and GOF at NASA's GSFC for provision of the ASM data.  LH acknowledges support of a PPARC studentship.
\end{ack}






\bibliographystyle{myjphyB2}
\bibliography{m15}
}
\end{document}